# Physical Foundation for General Interior Tomography

Ge Wang, PhD, Rensselaer Polytechnic Institute, Troy, NY, USA

ge-wang@ieee.org

December 15, 2014

**Abstract**

Gauge invariability guarantees the same form of Maxwell's equations in different coordinate systems, and is instrumental for electromagnetic cloaking to hide a region of interest (ROI) perfectly. On the other hand, interior tomography is to reconstruct an ROI exactly. In this article, the recent results in these two disconnected areas are brought together to justify the general interior tomography principle. Several opportunities are suggested for tomographic research.

## I. Introduction

Visibility and invisibility are opposite concepts but in a tomographic imaging perspective these concepts are intimately related. While superficial views of daily objects are most common, cross-sectional images are routinely examined in clinics and hospitals with tomographic imaging modalities especially x-ray CT. On the other hand, a cloak of invisibility has been a hot research topic, whose utility is to shield an object from physical probing such as electromagnetic radiation and ultrasonic scanning. Without loss of generality, in this article the main focus is placed on the relevance between interior tomography and electromagnetic cloaking – The former is to achieve an inner vision over a region of interest (ROI) in a theoretically exact fashion, and the latter is to make the ROI virtually transparent to an electromagnetic field.

In the next section, interior tomography is explained in reference to (Wang and Yu 2013). In the third section, electromagnetic cloaking is described based on (Pendry, Schurig et al. 2006). In the fourth section, the argument for electromagnetic cloaking is applied to establish a broad physical foundation for the general interior tomography principle. In the last section, further topics are discussed.

## II. Interior Tomography

Radiographic imaging only gives a projective view. X-ray CT collects many projections. Mathematically, what can be measured with pencil x-ray beams are line integrals. Performing a 1D Fourier transformation of a parallel-beam projection profile, we obtain a profile along a radial line in the 2D Fourier space. Varying the projection orientation by an angle, the direction of the radial line is changed by the same angle in the Fourier space, and the whole 2D Fourier space can be fully covered during a full scan. Then, the original image can be exactly reconstructed using an inverse Fourier transform. CT is nothing but image reconstruction from projections.

Parallel-beam geometry is not efficient. Fan-beam geometry is much faster. To collect a sufficient amount of fan-beam data for accurate and stable image reconstruction of a whole cross section, it is requested that there is at least a source point on any line through an object to be reconstructed. Even more efficient is cone-beam geometry. To collect enough cone-beam data for accurate and stable reconstruction of an entire object, it is requested that there is at least a source point on any plane through an object.

The classic CT strategy or the central dogma in the CT field has been that a whole cross-section or entire object must be completely covered by a fan- or cone-beam of x-rays, resultant data must be explicitly or implicitly transformed to the Radon (line or planar integrals) or Fourier space that must be fully sampled for accurate and stable image reconstruction.

In contrast to the classic dogma that Radon data must be complete for global image reconstruction, the modern reconstruction research has been largely motivated to handle data truncation so that image reconstruction can still be theoretically exact. The cone-beam spiral CT reconstruction is the longitudinal data truncation problem. The second generation of CT scanning uses a narrow fan-beam. The translation of the x-ray source is to avoid transverse data truncation problem, which is also known as the interior problem. In the long-standing interior problem, only x-rays that go through an interior region of interest (ROI) are used which generate truncated projection data. This interior problem was proven no unique solution more than a decade ago. Since 2007, this problem has been re-investigated by several groups (Wang and Yu 2013). In 2007, it was proved that if there is a known sub-region in the ROI the interior problem can be uniquely and stably solved (Ye, Yu et al. 2007). The assumption on a known sub-region is practical; for example, air in an airway, blood in an aorta, or an intact feature from a prior scan. However, once a contrast agent is injected for contrast-enhanced cardiac or cancer studies, blood density cannot be assumed as known. From 2009 to 2010, it was proved that if an ROI can be decomposed into piecewise constant/polynomial sub-regions, the solution to the interior problem is unique (Han, Yu et al. 2009, Yu and Wang 2009, Yu, Yang et al. 2009, Yang, Yu et al. 2010). In 2012, it was further proved that for a piecewise polynomial ROI the solution to the interior problem is not only unique but also stable (Katsevich, Katsevich et al. 2012). Interior tomography is not just a mathematical curiosity, and it produces excellent local reconstruction in practice (Yu and Wang 2009, Wang and Yu 2010, Yu, Wang et al. 2011).

Excluding non-local data from a global scan and yet achieving theoretically exact ROI reconstruction is a breakthrough in the field of x-ray CT. The implication of such data decomposition is beyond this particular imaging modality (Wang, Zhang et al. 2012). Indeed, interior tomography has been generalized from x-ray CT to SPECT (Yu, Yang et al. 2009, Yang, Yu et al. 2012), MRI (Wang, Zhang et al. 2012), and phase-contrast tomography (Cong, Yang et al. 2012, Yang, Cong et al. 2012). The general interior tomography principle suggested in (Wang, Zhang et al. 2012, Wang and Yu 2013) can be described as follows:

> *Tomographic imaging of an interior region of interest (ROI) can be in principle exactly and stably performed in an appropriate space (such as a family of piecewise polynomial functions) from an interior dataset **I** of a global dataset **G** where **G** contains indirectly measured data sufficient for theoretically exact and stable reconstruction over the support of a whole cross section or entire object, and **I** contains and only contains these indirectly measured data that directly involve the ROI.*

This principle can have an alternative description that *localized tomographic reconstruction needs and only needs local data*. In other words, *tomographic characterization of an interior ROI can be performed with the least amount of information in the form of an interior scan*. This vision suggests major research opportunities. In a recent review paper on interior tomography (Wang and Yu 2013), a mathematical hypothesis on general interior tomography is presented that

> *"Although interior tomography theory has been developed based on individual imaging modalities (e.g. CT, SPECT, phase-contrast tomography), we believe that the interior tomography principle can be refined for a rigorous unification. Let us consider a general weighted integral over a sub-domain of an object, where the sub-domain is compact with a smooth boundary, and controlled by two parameters. When the first parameter of the subdomain is fixed, varying the second parameter will move the sub-domain smoothly over an ROI in one direction specified by the first parameter. Specifically, let us assume that the measurement be in the form of a **P** transform of weighted integrals over sub-domains, where **P** is a polynomial of differential operators up to a finite order. It is our hypothesis that theoretically exact and stable ROI reconstruction can be made from the generalized measures that directly involve an ROI; that is, (1) the intersection of each sub-domain and the ROI is non-empty, (2) all measures satisfying (1) are available, and (3) the ROI is sparse satisfying the piecewise constant/polynomial model or in another linear transform domain."*

### III.     Electromagnetic Cloaking

A cloak of invisibility used to be a fictional object but it has now been a device under technical development based on some of the finest physical theories. In this subsection, the theory for controlling electromagnetic fields is briefly developed, which was first presented Pendry, Schurig, and Smith in *Science*, 2006 (Pendry, Schurig et al. 2006). Their work demonstrates that a perfect cloaking of an object in an electromagnetic field is theoretically possible and practically feasible.

Their study assumes metamaterials whose novel properties were reported for a wide portion of the electromagnetic spectrum (Dolling, Enkrich et al. 2006, Schurig, Mock et al. 2006, Valentine, Zhang et al. 2008, Billings 2013, Hunt, Driscoll et al. 2013). As a result, it is in principle feasible that a material can be designed with arbitrary distributions of permittivity and permeability values (positive or negative) specified independently. By modifying electromagnetic properties of an inhomogeneous composite metamaterial, any reasonable electromagnetic field can be formed relative to a homogeneous reference field, with "*the conserved quantities of electromagnetism — the electric displacement field **D**, the magnetic field intensity **B**, and the Poynting vector **S** — can all be directed at will*" (Pendry, Schurig et al. 2006). This is a result of the profound gauge invariability that governs Maxwell's equations, and more.

Electromagnetic cloaking is a special application of the above theoretical observation. With a metamaterial, the electromagnetic field lines can be smoothly deformed to go around an ROI and then back to their original trajectories as if the particular ROI is in non-existence. This is physically realistic directly based on

Maxwell's equations without any approximation. In other words, when the electromagnetic field lines are distorted, Maxwell's equations have exactly the same form in the new coordinate system except that both the permittivity and permeability are scaled by a common factor (Pendry, Schurig et al. 2006). Therefore, there is a total freedom in controlling an electromagnetic field, as shown in Figure 1.

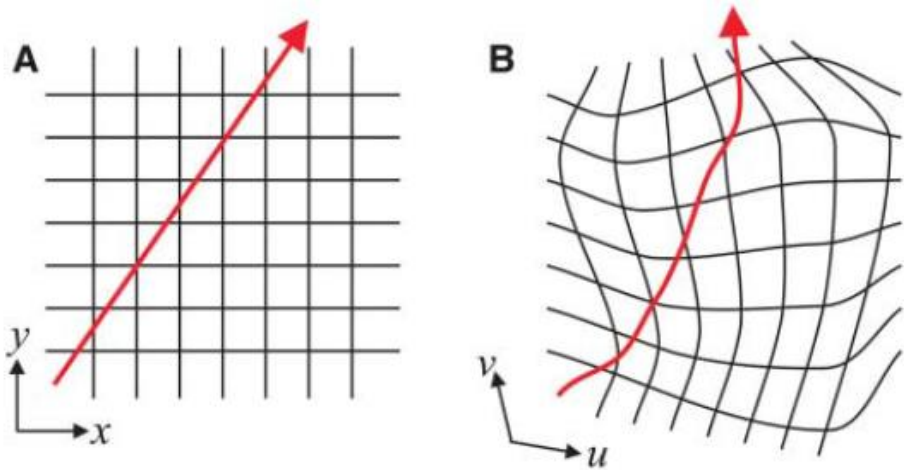

Fig. 1. (**A**) A field line in free space with the background Cartesian coordinate grid shown. (**B**) The distorted field line with the background coordinates distorted in the same fashion. The field in question may be the electric displacement or magnetic induction fields **D** or **B**, or the Poynting vector **S**, which is equivalent to a ray of light (Copied from (Pendry, Schurig et al. 2006)).

### IV. Physical Insight for General Interior Tomography

In this subsection, the above scheme for electromagnetic field deformation is applied as a machinery to generate a large family of *physically feasible interior tomography settings*, among which x-ray interior tomography is only a special case. In this context, it becomes clear and inspiring under what conditions which new forms of interior scans can be performed as for x-ray interior tomography. To present this physical insight, Figure 2(A) and 2(B) shows corresponding interior scans in the homogeneous and inhomogeneous imaging geometries respectively, and the diffractive parameters are known in either case.

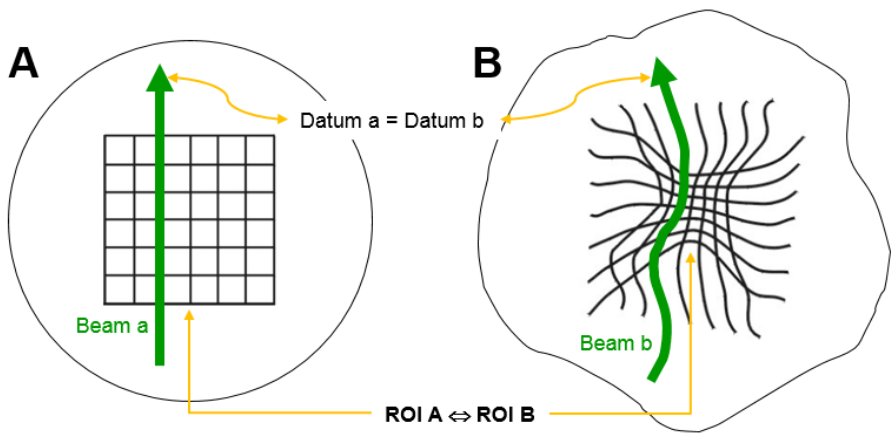

Fig. 2. (**A**) Interior tomography in the reference framework where a typical ray a goes through an ROI A to produce a datum a. (**B**) A geometrically deformed and yet physically feasible interior tomography problem where a typical probing beam is curvilinear and has a varying width. The same amount of information is collected in the cases **A** and **B** respectively on the equivalent ROIs **A** and **B**.

If the straight green line in Figure 2(A) is interpreted as an x-ray pencil beam which has a certain width, the corresponding curvilinear path in Figure 2(B) would have varying widths, which is physically feasible and delivers the equivalent measurement as that recorded in Figure 2(A). This observation is the key to generalize x-ray interior tomography to a general imaging geometry. Specifically, through an ROI in Figure 2(A) truncated projective measurements can be performed along straight paths through the ROI. Equivalently, the same amount of information can be collected along bundles of curvilinear paths through the corresponding deformed ROI in Figure 2(B). In fact, the data come from longitudinal manifolds in Figure 2(B) are equivalent to a truncated x-ray projection along the vertical direction in Figure 2(A). Therefore, as long as sufficient information for tomographic reconstruction is available for exact image reconstruction over the ROI in Figure 2(A), the corresponding measurements in Figure 2(B) ought to be sufficient for exact image reconstruction over the deformed ROI in Figure 2(B), despite that the data are transversely truncated in both Figure 2(A) and 2(B).

It is underlined that when multiple projective measurements are involved over an ROI, the deformed measurement geometry is not necessarily the same from projection to projection. That is, the deformed field shown in Figure 2(B) can be different for each projective measurement. Nevertheless, each measurement in Figure 2(B) can be related to a projective measurement in Figure 2(A). Ultimately, it is required that the interior dataset in Figure 2(A) should support an accurate interior reconstruction. Then, the interior reconstruction over the ROI in Figure 2(B) can be done from the corresponding dataset collected in Figure 2(B), to a degree of accuracy and precision consistent to that for Figure 2(A).

## V.    Discussions & Conclusion

The electromagnetic mechanism covers not only "particles" of different energies in a beam but also "waves" of various frequencies. In a wave perspective, the blue wave front in Figure 3(A) can be deformed into that in Figure 3(B). In this way, current phase-contrast and dark-field imaging methods can be in principle changed into new forms: from a homogeneous background such as air or water into an inhomogeneous background made of metamaterial. It is conceivable that a liquid/gas-like metamaterial background can be dynamically controlled to emulate a tomographic scan when an object remains embedded in such a medium, which is not practical at this moment yet.

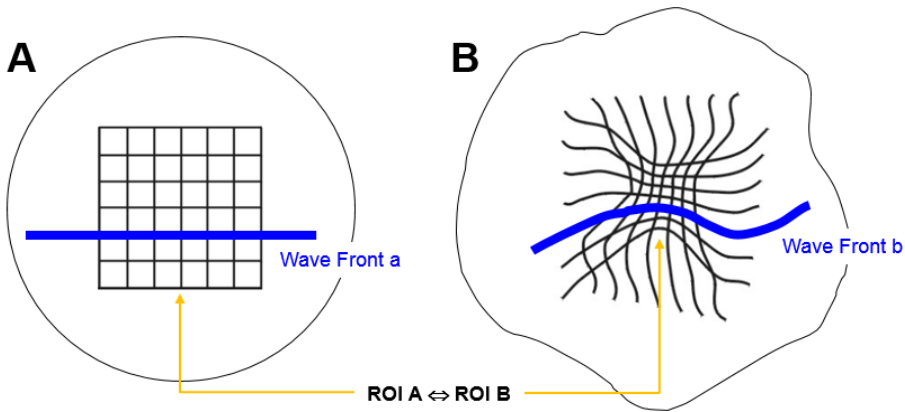

Fig. 3. (**A**) Wave tomography in the reference framework where a wave front propagates through an ROI A. (**B**) A geometrically deformed wave front that is physically feasible in the inhomogeneous medium.

In mathematical terms, a physical field deformation is an isomorphism, which admits a unique inverse. That is, two images of an ROI are isomorphic before and after the field deformation. Moreover, measurements on the ROI in one image is equivalent to the corresponding data for the other image. These measurement processes are in inter-related fields but different coordinate systems, and yet still obey the same governing equations *in the same mathematical form*. In this regard, differential geometry and field theory seem very useful for further investigation of general interior tomography as a new area of integral geometry.

The main point of this article is to argue for the existence of a wide class of electromagnetically meaningful interior tomography problems that are in continuously deformed imaging geometries. This physical foundation for general interior tomography has its root in the famous gauge invariability. Given the generality of the gauge invariability, it is hypothesized that further physics-based extension of tomographic imaging geometries and methods can be pursued, such as for an ultrasonic wave field. It is also noted that global tomography can be treated as a special case of interior tomography (the degree of data truncation is zero).

In conclusion, a physical insight on general interior tomography has been brainstormed in light of the recent work on electromagnetic cloaking. This seems a solid step towards a comprehensive physical foundation for a unified tomographic imaging theory, especially in restricted scenarios due to data truncation and noise. An interesting topic is the extension of interior tomography to higher dimensionality, which may lead to interesting results in even and odd dimensions.

## References

1. Billings, L. (2013). "Metamaterial World." Nature 500(7461): 138-140.
2. Cong, W. X., J. S. Yang and G. Wang (2012). "Differential phase-contrast interior tomography." Physics in Medicine and Biology 57(10): 2905-2914.
3. Dolling, G., C. Enkrich, M. Wegener, C. M. Soukoulis and S. Linden (2006). "Simultaneous negative phase and group velocity of light in a metamaterial." Science 312(5775): 892-894.
4. Han, W., H. Yu and G. Wang (2009). "A total variation minimization theorem for compressed sensing based tomography." International Journal of Biomedical Imaging 2009: Articel ID:125871, 125873 pages.
5. Hunt, J., T. Driscoll, A. Mrozack, G. Lipworth, M. Reynolds, D. Brady and D. R. Smith (2013). "Metamaterial Apertures for Computational Imaging." Science 339(6117): 310-313.
6. Katsevich, E., A. Katsevich and G. Wang (2012). "Stability of the interior problem with polynomial attenuation in the region of interest." Inverse Problems 28(6).
7. Pendry, J. B., D. Schurig and D. R. Smith (2006). "Controlling electromagnetic fields." Science 312(5781): 1780-1782.
8. Schurig, D., J. J. Mock, B. J. Justice, S. A. Cummer, J. B. Pendry, A. F. Starr and D. R. Smith (2006). "Metamaterial electromagnetic cloak at microwave frequencies." Science 314(5801): 977-980.
9. Valentine, J., S. Zhang, T. Zentgraf, E. Ulin-Avila, D. A. Genov, G. Bartal and X. Zhang (2008). "Three-dimensional optical metamaterial with a negative refractive index." Nature 455(7211): 376-U332.
10. Wang, G. and H. Y. Yu (2010). "Can interior tomography outperform lambda tomography?" Proceedings of the National Academy of Sciences of the United States of America 107(22): E92-E93.
11. Wang, G. and H. Y. Yu (2013). "The meaning of interior tomography." Physics in Medicine and Biology 58(16): R161-R186.
12. Wang, G., J. Zhang, H. Gao, V. Weir, H. Y. Yu, W. X. Cong, X. C. Xu, H. O. Shen, J. Bennett, M. Furth, Y. Wang and M. Vannier (2012). "Towards Omni-Tomography-Grand Fusion of Multiple Modalities for Simultaneous Interior Tomography." Plos One 7(6).
13. Yang, J. S., W. X. Cong, M. Jiang and G. Wang (2012). "Theoretical study on high order interior tomography " Journal of X-ray Science and Technology 20: 423-436.
14. Yang, J. S., H. Y. Yu, M. Jiang and G. Wang (2010). "High-order total variation minimization for interior tomography." Inverse Problems 26(3).
15. Yang, J. S., H. Y. Yu, M. Jiang and G. Wang (2012). "High-order total variation minimization for interior SPECT." Inverse Problems 28(1).
16. Ye, Y., H. Yu, Y. Wei and G. Wang (2007). "A general local reconstruction approach based on a truncated Hilbert transform." International Journal of Biomedical Imaging 2007: Article ID: 63634, 63638 pages.
17. Yu, H. Y. and G. Wang (2009). "Compressed sensing based interior tomography." Physics in Medicine and Biology 54(9): 2791-2805.
18. Yu, H. Y., G. Wang, J. Hsieh, D. W. Entrikin, S. Ellis, B. D. Liu and J. J. Carr (2011). "Compressive Sensing-Based Interior Tomography: Preliminary Clinical Application." Journal of Computer Assisted Tomography 35(6): 762-764.
19. Yu, H. Y., J. S. Yang, M. Jiang and G. Wang (2009). "Interior SPECT-exact and stable ROI reconstruction from uniformly attenuated local projections." Communications in Numerical Methods in Engineering 25(6): 693-710.
20. Yu, H. Y., J. S. Yang, M. Jiang and G. Wang (2009). "Supplemental analysis on compressed sensing based interior tomography." Physics in Medicine and Biology 54(18): N425-N432.